\documentclass[pre,aps,twocolumn,showpacs,floatfix,superscriptaddress]{revtex4-2}
\usepackage{graphicx}
\usepackage{bm}
\usepackage{color}
\usepackage{amssymb,amsmath}
\usepackage{ulem}
\usepackage{ragged2e}
\usepackage{adjustbox}
\usepackage{multirow}
\usepackage{booktabs}
\usepackage[table]{xcolor}
\usepackage{tcolorbox}
\usepackage{tabularx}
\usepackage{array}
\usepackage{placeins}
\usepackage{colortbl}
\usepackage{hyperref}
\hypersetup{colorlinks=true, citecolor=blue, urlcolor=blue, linkcolor=blue}
\tcbuselibrary{skins}
\definecolor{Salmon}{RGB}{250,128,114}

\newcolumntype{Y}{>{\raggedleft\arraybackslash}X}

\tcbset{tab1/.style={\centering, fonttitle=\bfseries\small,fontupper=\normalsize\sffamily,
		colback=yellow!10!white,colframe=red!75!black,colbacktitle=Salmon!40!white,
		coltitle=black,center title,freelance,frame code={
			\foreach \n in {north east,north west,south east,south west}
			{\path [fill=red!75!black] (interior.\n) circle (3mm); };},}}

\tcbset{tab2/.style={enhanced,fonttitle=\bfseries,fontupper=\small\sffamily,
		colback=yellow!10!white,colframe=red!50!black,colbacktitle=Salmon!40!white,
		coltitle=black,center title}}
\begin{document}	
	\title{Delay-Controlled Heterogeneous Nucleation in Adaptive Dynamical Networks}
	\author{R Anand}
	\affiliation{Centre for Nonlinear Science and Engineering, Department of Physics, School of EEE, SASTRA Deemed to be University, Thanjavur, 613 401. }
	
	\author{Jan Fialkowski}
	\affiliation{
		Complexity Science Hub, Metternichgasse 8, A-1030, Vienna, Austria		
	}
	\affiliation{	Institute of the Science of Complex Systems, CeDAS, Medical University Vienna, Spitalgasse 23, A-1090, Vienna, Austria}
	\author{V K Chandrasekar}
	\affiliation{Centre for Nonlinear Science and Engineering, Department of Physics, School of EEE, SASTRA Deemed to be University, Thanjavur, 613 401. }
	\author{R Suresh}
	\email{suresh@eee.sastra.edu}
	\affiliation{Centre for Nonlinear Science and Engineering, Department of Physics, School of EEE, SASTRA Deemed to be University, Thanjavur, 613 401. }

\begin{abstract}
	Phase transitions constitute fundamental mechanisms underlying abrupt or qualitative changes in the collective dynamics of interacting units across a wide range of natural and engineered systems. In dynamical networks, such transitions lead to significant reorganization in the coordinated behavior of coupled elements. In adaptive dynamical networks, the connectivity evolves dynamically in response to the states of the nodes, resulting in a coevolution of structure and dynamics. In this work, we report two distinct forms of heterogeneous nucleation that give rise to single-step and multi-step phase transitions toward global synchronization in finite-size adaptive networks with connection delays. We demonstrate that the nature of the nucleation transition is governed by both the presence and magnitude of the delay, as well as the class of natural frequency distribution. Using a collective coordinate framework, we develop a mean-field description of cluster dynamics and derive an analytical upper bound condition for the existence of two-cluster states, which shows excellent agreement with numerical simulations. Furthermore, we extend the analysis to systems with distributed delays and obtain corresponding analytical conditions. Our results provide a theoretical framework for understanding synchronization transitions in adaptive networks with time-delayed interactions.
\end{abstract}
\maketitle	
\section{Introduction}
Synchronization transitions represent a fundamental collective phenomenon that plays a central role across a wide range of disciplines, including physics~\cite{wiesenfeld1996synchronization}, chemistry~\cite{taylor2009dynamical}, biology~\cite{winfree1967biological}, and network science~\cite{arenas2008synchronization}. Complex networks, in particular, exhibit a rich variety of nonequilibrium phase transitions that govern their collective dynamics in response to variations in control parameters such as coupling strength, interaction delay, and noise intensity~\cite{stanley1971phase,newman2003structure,boccaletti2018synchronization,scholl2012nonequilibrium}. 

Such transitions are well documented in classical physical systems, including crystallization~\cite{guo2016kinetic} and ferromagnetic ordering, and have direct analogues in cooperative behaviors observed in biological and engineered systems~\cite{gambuzza2020controlling}. The nature of synchronization transitions in a network is primarily determined by two key factors: (i) the distribution of natural frequencies of the oscillators and (ii) the underlying network topology, including properties such as degree distribution, clustering, modularity, and spatial structure~\cite{xu2019universal,rodrigues2016kuramoto}. These factors collectively determine whether the transition is continuous, abrupt, multistep, or characterized by intermediate partially synchronized states such as frequency clusters~\cite{menara2019stability}, chimera states~\cite{frolov2020chimera}, or traveling waves~\cite{luccon2017long}.

In recent years, adaptive dynamical networks have attracted significant attention due to their ability to capture the coevolution of network topology and node dynamics~\cite{abbott2000synaptic,berner2023adaptive,taylor2010spontaneous,horstmeyer2020adaptive,markram1997regulation,jain2001model}. These systems exhibit a wide range of emergent phenomena, including desynchronization transitions~\cite{berner2021desynchronization}, partial synchronization patterns, and heterogeneous nucleation driven by frequency disorder~\cite{fialkowski2023heterogeneous}. The adaptive nature of these networks introduces feedback mechanisms between structure and dynamics, enabling self-organized synchronization and complex collective behavior\cite{aoki2009co,aoki2011self,kasatkin2017self,kasatkin2018synchronization,berner2019hierarchical,berner2020birth}

When the adaptation timescale is much slower than the intrinsic oscillator dynamics, singular perturbation theory can be employed to understand how slow plasticity shapes emergent states~\cite{berner2023adaptive}. In biological systems, adaptation rules are often inspired by synaptic plasticity mechanisms such as Hebbian learning and spike-timing-dependent plasticity (STDP)~\cite{gerstner1998neuronal,caporale2008spike,bi2001synaptic,meisel2009adaptive,lucken2016noise}. These rules strengthen connections between coherently firing neurons while weakening incoherent interactions~\cite{donald1949organization,lowel1992selection,zhang1998critical}.

Incorporating additional physical effects such as inertia, delays, and heterogeneous interactions has been shown to further enrich synchronization dynamics~\cite{berner2021adaptive,anand2026emergence,yadav2024disparity,yadav2025heterogeneous,senthamizhan2025swarmalators,xu2016synchronization}. In particular, time delays play a crucial role in realistic systems, as signal transmission is inherently non-instantaneous~\cite{timms2014synchronization,thamizharasan2025dynamics,thamizharasan2024stimulus}. Understanding how delays interact with adaptive mechanisms is therefore essential for describing the dynamics of real-world networks.

In this work, we investigate heterogeneous nucleation in adaptive dynamical networks in the presence of connection delays. We demonstrate that delays can control the transition between distinct nucleation pathways in the adaptive Kuramoto model. While previous studies have reported both single-step and multi-step transitions depending on frequency disorder~\cite{fialkowski2023heterogeneous}, those results often relied on more complex frameworks such as multiplex or multi-population networks~\cite{yadav2024disparity,yadav2025heterogeneous}. In those cases, the nature of the transition depends strongly on the adaptation rate disparity: interpopulation adaptation rate tends to favor multi-step transitions, whereas intrapopulation adaptation rate typically leads to single-step transitions across all different classes of frequency disorders. However, those results were obtained in relatively more complex modeling frameworks involving additional population structure or network layers.

Here, we show that heterogeneous nucleation can arise in a simpler single-population adaptive Kuramoto model when connection delays are introduced. The delay induces an effective phase lag that depends on cluster frequencies, thereby altering the transition pathway in a manner dependent on the frequency distribution.

To systematically investigate this effect, we consider two classes of natural frequency distributions: (i) class-I distributions with a strong concentration around the mean frequency and (ii) class-II distributions with bimodal structure. We demonstrate that connection delay can induce a transition between single-step and multi-step synchronization pathways in both cases.

The remainder of the paper is organized as follows. In Sec.~\ref{model}, we introduce the model. Section~\ref{results} presents numerical results and characterizes the nucleation transitions. In Sec.~\ref{analytical}, we develop a reduced mean-field description and derive analytical conditions. Section~\ref{dis} extends the analysis to distributed delays. Finally, Sec.~\ref{conclusion} summarizes the main findings.

\section{Model}\label{model}
We consider the adaptive Kuramoto model, a widely used framework for studying synchronization and emergent behavior in complex dynamical systems. An important extension of this model involves incorporating time delays, which render the effective coupling between oscillators asymmetric~\cite{timms2014synchronization,yeung1999time,wu2018dynamics}.

In neuronal systems, for instance, signals are processed based on their arrival times rather than their emission times~\cite{stam2014modern}. This makes the inclusion of communication delays essential for realistic modeling.

A standard approach to incorporating delay is to replace the instantaneous phase difference in the coupling term with a delayed phase difference. The resulting delay-coupled adaptive Kuramoto model is given by

\begin{subequations}
	\label{eq1}
	\begin{eqnarray}
		\label{eq1a}
		\dot\phi_i(t) &=& \omega_i - \frac{\sigma}{N} \sum_{j=1}^{N} k_{ij} \sin(\phi_i(t) - \phi_j(t - \tau)),
	\end{eqnarray}
	\begin{eqnarray}
		\label{eq1b}
		\dot k_{ij}(t) &=& -\epsilon \; (k_{ij} + \sin(\phi_i(t) - \phi_j(t - \tau) + \beta)),
	\end{eqnarray}
\end{subequations}

where $\phi_i$ denotes the phase of the $i$-th oscillator and $k_{ij}$ represents the adaptive coupling weight between oscillators $i$ and $j$. The natural frequencies $\omega_i$ are drawn in the interval $\omega_i \in [-0.25, 0.25]$.

The parameter $\sigma$ denotes the global coupling strength and serves as a control parameter, $\epsilon$ is the timescale separation parameter $(\epsilon = 0.01)$, $\tau$ is the connection delay and serves as a second control parameter, and $\beta$ is the adaptation control parameter. Throughout this work, we fix $\beta = -0.53\pi$, corresponding to symmetric Hebbian-like adaptation.

\section{Results} \label{results}
The phase oscillators equation.~(\ref{eq1}) is numerically integrated using a fourth-order Runge–Kutta algorithm with a fixed time step of $\Delta t = 0.01$. Here, we choose the network size $N = 50$. The initial phases $\phi_i(0)$ are drawn randomly from a uniform distribution in the interval $[0, 2\pi)$, while the initial coupling weights are $k_{ij}(0) = 0 \;\forall\; i, j$. The parameters are fixed at $\epsilon = 0.01$ and $\beta = -0.53\pi$. 

Next, we employ the measure for coherence of the frequency synchronized oscillator pairs, it can be quantified by the synchronization index $S$ \cite{berner2021desynchronization,fialkowski2023heterogeneous}, which is defined as 
\begin{eqnarray}
	\label{S}
	S = \frac{1}{N^2} \sum_{i = 1}^{N} \sum_{j = 1}^{N} s_{ij},
\end{eqnarray} 
where, $s_{ij}$ is the frequency synchronization measure between nodes and is given by
\begin{eqnarray}
	\label{sij}
	s_{ij} = \begin{cases}
		1, & \text{if }  \lvert \langle \dot{\phi}_i \rangle -\langle \dot{\phi}_j \rangle\rvert \leq \delta, \\
		0, & \text{otherwise},
	\end{cases}
\end{eqnarray}
where, $\delta$ is the predefined threshold value and $\langle \dot{\phi}_i \rangle$ is the time-averaged frequency (mean-phase velocity). Here, $\delta = 0.001$ is a predefined threshold for the difference in the time-averaged frequency between the $i$-th and $j$-th nodes. The time-averaged frequency $\langle \dot{\phi}_i \rangle$ for sufficient transient time $T_0$ is defined as 
\begin{eqnarray}
	\label{freq}
	\langle \dot{\phi}_i \rangle &=& \lim_{T \rightarrow \infty} \frac{1}{T} \int_{T_0}^{T_0 + T} \dot{\phi}_i(t) dt, \\ 
	&=&\lim_{T \rightarrow \infty} \frac{1}{T} (\phi_i(T_0 + T) - \phi_i(T_0)) \nonumber.
\end{eqnarray}

For $S = 1$ the system is completely frequency synchronized, while $S = 0$ the system is desynchronized. As the coupling strength $ \sigma  $ is increased, the system undergoes a transition from a desynchronized to a fully frequency-synchronized state.

In the following subsections, we explain how delay influences the routes to synchrony with increasing $  \sigma  $. Specifically, we show that the nature of the transition can follow two distinct first-order paths: a single-step transition or a multi-step transition, depending on the delay.
\begin{figure}[!ht]
	\centering
	\includegraphics[width=0.5\textwidth]{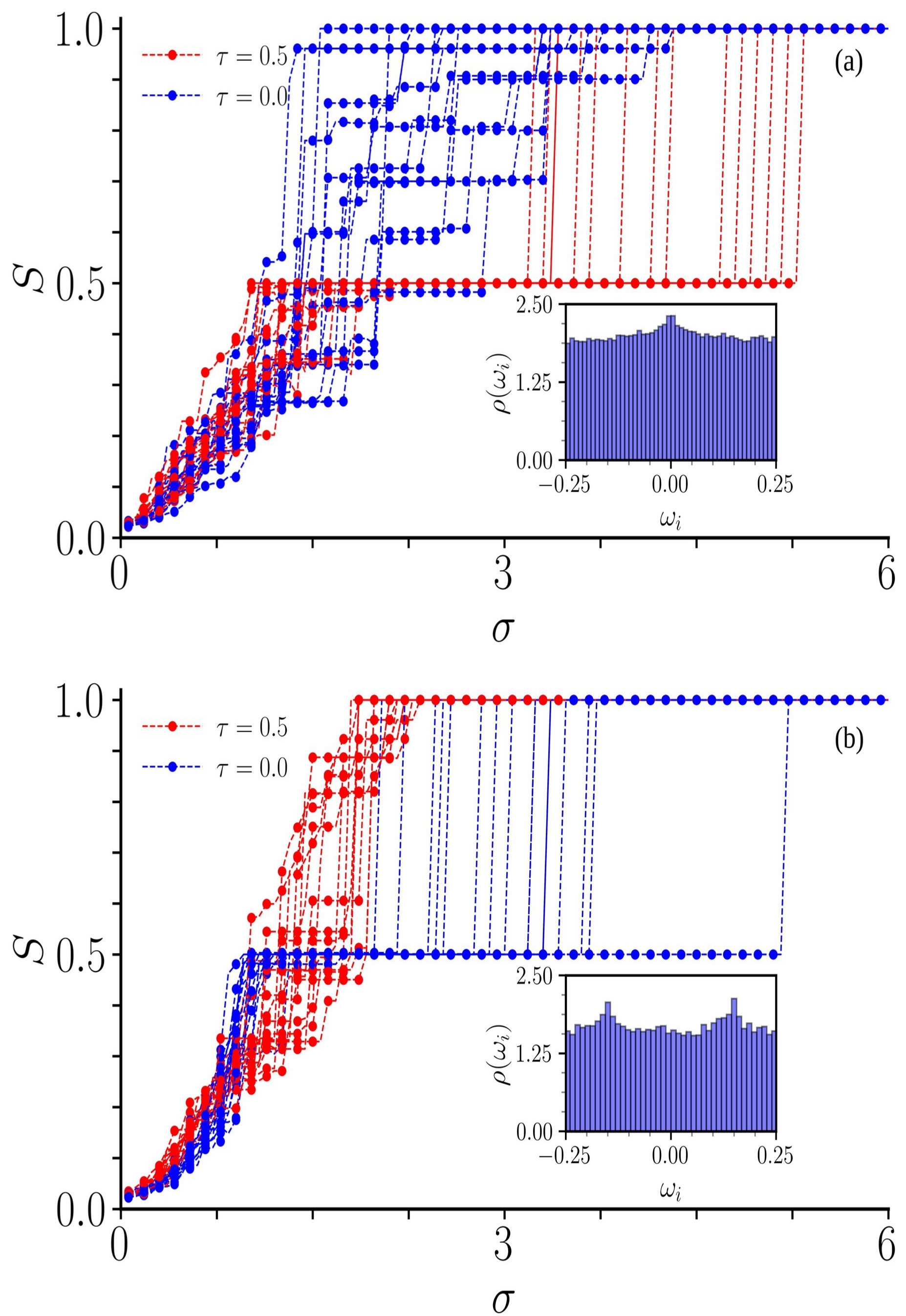}
	\caption{Synchronization index $S$ as a function of coupling strength $\sigma$ obtained from $200$ independent simulations. Results are shown for (a) class-I and (b) class-II natural frequency distributions, both in the absence and presence of delay. Red filled circles indicate presence of delay, whereas blue filled circles denote absence of delay. The insets display representative realizations of the corresponding natural frequency distributions. The parameters are $\beta = -0.53\pi$, $\epsilon = 0.01$, and $\delta = 0.001$.}
	\label{1p_num}
\end{figure}
\subsection{Class-I natural frequency distribution}
We first examine the influence of connection delay on synchronization transitions for class-I frequency distributions, characterized by a strong concentration of oscillators around the mean frequency.

Figure~\ref{1p_num}(a) shows the synchronization index $S$ as a function of the coupling strength $\sigma$. In the absence of delay ($\tau = 0$), the system exhibits a clear multi-step transition to global synchronization (blue filled circles). In contrast, introducing a finite delay ($\tau = 0.5$) results in a single-step (abrupt) transition (red filled circles).

This behavior can be understood in terms of heterogeneous nucleation. For class-I distributions, the high density of oscillators near the mean frequency promotes the formation of an initial synchronized nucleus. As the coupling strength increases, this nucleus progressively entrains oscillators with increasingly different natural frequencies, leading to a hierarchical growth process. Consequently, the order parameter evolves through multiple discrete steps, reflecting successive cluster absorption events as reported in \cite{fialkowski2023heterogeneous}.

To elucidate the mechanism by which delay alters this transition pathway, we consider an effective description of the adaptive coupling. Since the coupling weights evolve on a slow timescale ($\epsilon \ll 1$), they can be approximated as quasi-static on the timescale of phase dynamics:
\begin{equation}
k_{ij} \approx -\sin\big(\phi_i(t) - \phi_j(t-\tau) + \beta\big).
\label{eq5}
\end{equation}
Substituting this expression into the phase equation reveals that delay not only introduces memory effects but also fundamentally modifies the interaction structure. For oscillators belonging to a cluster with mean frequency $\Omega$, the delayed phase can be approximated as
\begin{equation}
\phi_j(t-\tau) \approx \phi_j(t) - \Omega \tau.
\label{eq6}
\end{equation}
This leads to an effective phase lag
\begin{equation}
\beta_{\text{eff}} = \beta + \Omega \tau,
\label{eq7}
\end{equation}
which depends explicitly on the cluster frequency.

This frequency-dependent phase lag constitutes the key mechanism underlying the delay-induced transition. In the absence of delay, all oscillators experience the same adaptation parameter $\beta$, and the transition is governed solely by the structure of the frequency distribution\cite{yeung1999time,klinshov2023kuramoto,montbrio2006time,skardal2018low}. As a result, synchronization proceeds via gradual, hierarchical cluster growth.

In contrast, when delay is present, different clusters experience distinct effective phase lags due to their differing frequencies. This breaks the uniformity of interactions and modifies the stability of multicluster configurations. When the delay-induced phase difference between clusters becomes sufficiently large, i.e., $|\Delta \Omega| \tau = \mathcal{O}(1)$, the system undergoes a qualitative reorganization, suppressing intermediate clustering and promoting a direct transition to global synchronization. Here 
$\Delta \Omega = (\Omega_\mu - \Omega_\nu)$, where $\mu$ and $\nu$ 
denote the cluster indices. In this regime, delay modifies the stability of multicluster configurations.

\begin{figure}[!ht]
	\centering
	\includegraphics[width=0.5\textwidth]{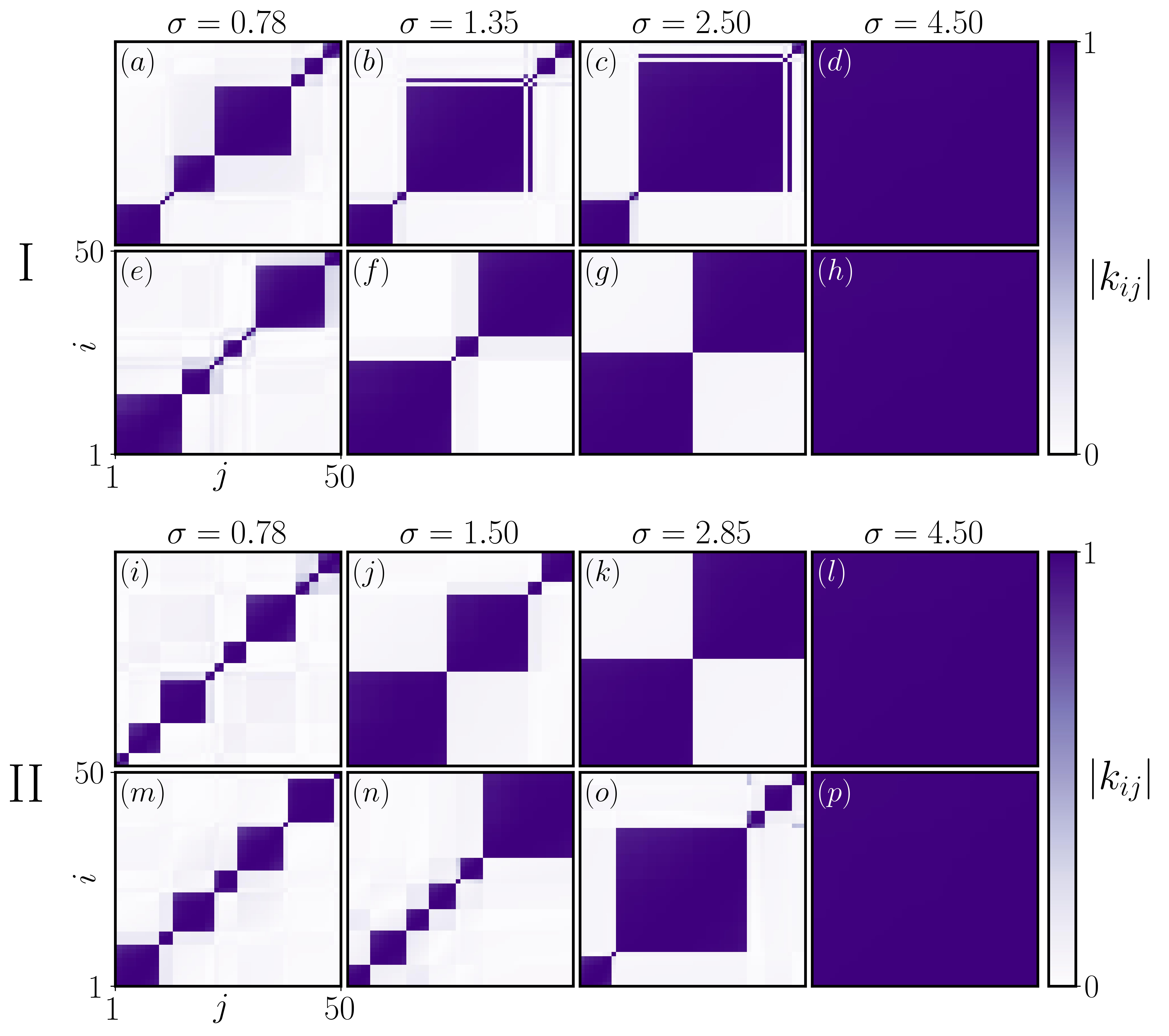}
	\caption{Evolution of the absolute value of the adaptive coupling matrix $|k_{ij}|$ with increasing coupling strength $\sigma$ for systems exhibiting heterogeneous nucleation. Panel I corresponds to class-I frequency distributions: the top row (a)–(d) shows multi-step transitions in the absence of delay ($\tau = 0$), while the bottom row (e)–(h) shows single-step transitions in the presence of delay ($\tau = 0.5$). Panel II corresponds to class-II frequency distributions: the top row (i)–(l) illustrates single-step transitions for $\tau = 0$, whereas the bottom row (m)–(p) shows multi-step transitions for $\tau = 0.5$. Each panel corresponds to increasing values of coupling strength $\sigma$. The parameters are $\beta = -0.53\pi$, $\epsilon = 0.01$, and $\delta = 0.001$.}
	\label{kij}
\end{figure}

The synchronization mechanism in the absence of delay is illustrated in Fig.\ref{kij}(a)–\ref{kij}(d), where we present the absolute value of the adaptive coupling between two oscillators $|k_{ij}|$. Blocks with a solid color correspond to a group of oscillators in a frequency cluster. Note that there is a single largest cluster that progressively grows by entraining the surrounding clusters, typical of multi-step transition. In contrast, in the presence of delay, $\tau = 0.5$, two equally sized clusters emerge, as shown in Fig.~\ref{kij}(e)–\ref{kij}(h). These two clusters synchronize for large values of of $\sigma = 0.78, 1.35, 2.50$, and $4.50$, giving rise to the single-step transition.

\subsection{Class-II natural frequency distribution}
We now consider class-II natural frequency distributions, characterized by two pronounced peaks away from the mean frequency. Such distributions naturally promote the formation of two dominant frequency clusters.

Figure~\ref{1p_num}(b) presents the synchronization index $S$ as a function of $\sigma$ for this case. In contrast to the class-I scenario, the system exhibits a single-step transition in the absence of delay ($\tau = 0$), whereas the introduction of delay ($\tau = 0.5$) leads to a multi-step transition.

At zero delay, the two symmetric clusters centered at $\pm \Omega$ experience identical interactions and remain stable until they abruptly merge into a globally synchronized state. This results in a discontinuous, single-step transition.

In this case, class-II frequency distributions, which are distinguished by a strong concentration of oscillators away from the mean frequency, promote nucleation at sites of frequency disorder. This leads to the formation of two fully entrained frequency clusters centered around these regions, resulting in single-step synchronization transitions in the absence of delay~\cite{fialkowski2023heterogeneous}.

The introduction of delay leads to the opposite behavior. While at zero delay the two symmetric clusters at $\pm \Omega$ experience identical interactions and coexist stably before merging abruptly, the presence of delay modifies this symmetry. In particular, the effective phase lags for the two clusters shift in opposite directions,
\begin{eqnarray}
	\beta_{\pm} \approx \beta \pm \Omega \tau.
	\label{eq8}
\end{eqnarray}
This delay-induced asymmetry breaks the balance between the clusters. Consequently, the previously stable two-cluster coexistence becomes unstable. Instead of collapsing through a single global synchronization event, the system evolves through a sequence of partial synchronization events, in which one cluster progressively entrains subsets of the other. This process gives rise to intermediate, partially synchronized states.

Importantly, the detailed synchronization pathway depends on finite-size fluctuations in the realization of the class-II natural frequencies. Small variations in the sampled peak densities and local frequency gaps influence the formation and evolution of cluster structures. These structural changes, in turn, feed back into the inter-cluster interactions through the adaptive coupling weights.

Because adaptive networks admit multiple self-consistent multicluster configurations for the same macroscopic parameters, such realization-dependent effects promote sequential cluster absorption and long-lived intermediate plateaus. As a result, the transition changes from a single-step abrupt synchronization to a multi-step synchronization pathway.

Figure~\ref{1p_num}(b) shows the synchronization index $S$ as a function of the coupling strength $\sigma$ for class-II frequency distributions. The single-step transition observed in the absence of delay ($\tau = 0$) is represented by blue filled circles, whereas the multi-step transition observed in the presence of delay ($\tau = 0.5$) is represented by red filled circles.

Further insight into the transition mechanism can be obtained from snapshots of the coupling weights $k_{ij}$ at increasing values of $\sigma$. Figures~\ref{kij}(i)–\ref{kij}(l) show the evolution of $|k_{ij}|$ for $\sigma = 0.78, 1.50, 2.85$, and $4.50$ in the absence of delay, illustrating the abrupt single-step transition. In contrast, Figs.~\ref{kij}(m)–\ref{kij}(p) correspond to the presence of delay ($\tau = 0.5$), where the system exhibits a clear multi-step transition for the same values of $\sigma$. We note that, for  $\tau = 0.5$, intermediate clusters may appear only temporarily and can disappear at later values of $\sigma$. Such cluster creation and subsequent reorganization is a typical feature of the delay-induced transition, where the effective phase-lag asymmetry destabilizes previously formed cluster configurations and promotes sequential restructuring of the multicluster state.

The next section presents a reduced model of the system and provides analytical estimates that support the numerical findings. The reduced description captures finite-size fluctuations inherent in systems of size $N$, while remaining consistent with the macroscopic behavior predicted by mean-field approximations.
\begin{figure}[!ht]	
	\centering
	\includegraphics[width=0.5\textwidth]{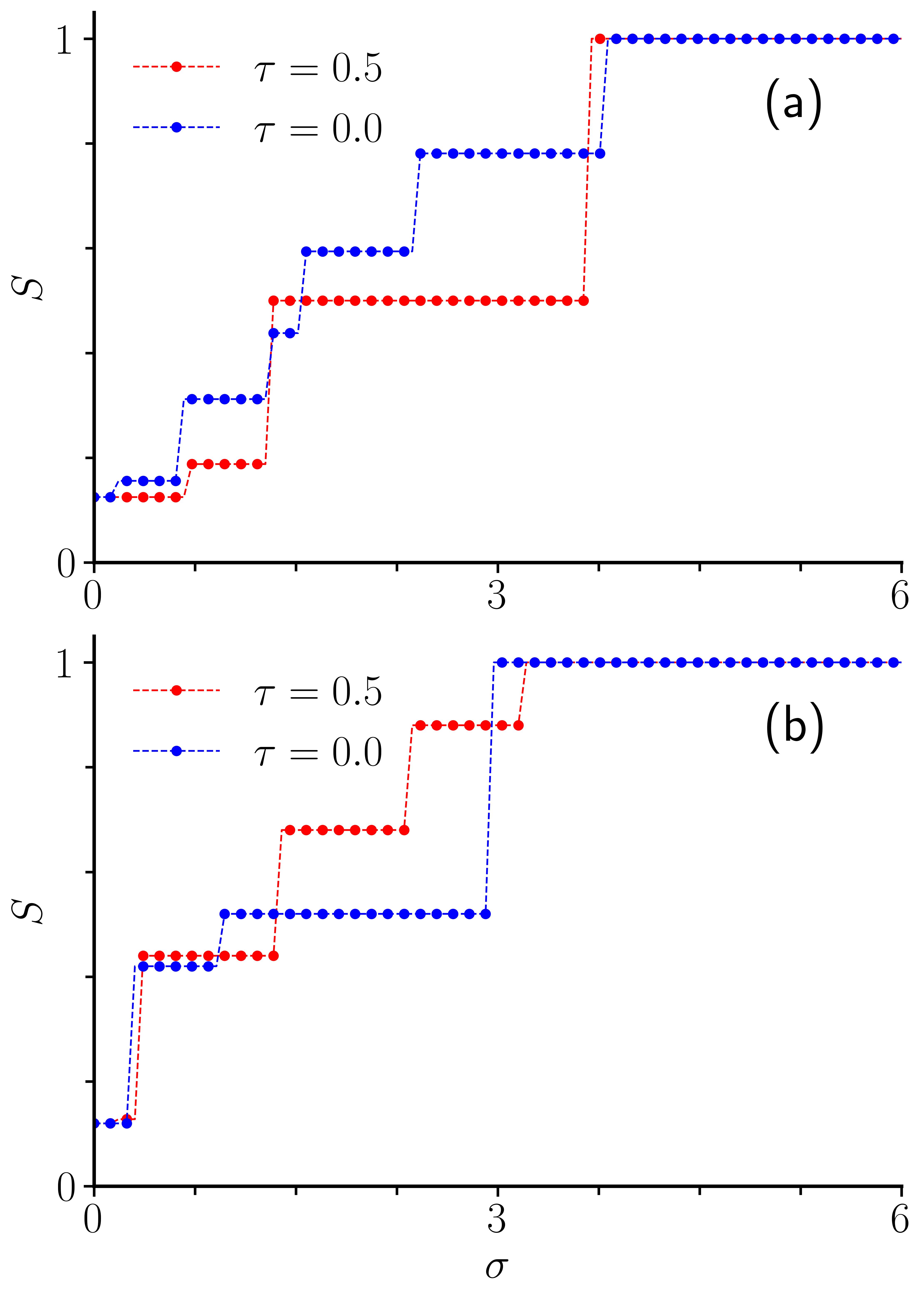}
	\caption{FIG. 3. Phase transitions obtained from the collective coordinate equations \eqref{an6} as a function of the coupling strength $\sigma$, corresponding to a prescribed six-cluster configuration. For panel (a), representing class-I frequency distributions, the cluster mean frequencies are chosen as $\Omega_\mu = (-0.225,\,-0.135,\,-0.045,\,0.045,\,0.135,\,0.225)$. For panel (b), representing class-II frequency distributions, the cluster mean frequencies are chosen as $\Omega_\mu = (-0.225,\,-0.185,\,-0.145,\,0.145,\,0.185,\,0.225)$. These values are selected such that they emulate the effective frequency organization observed in the full simulations. In panel (a), red filled circles denote the multi-step transition for $\tau = 0$, while blue filled circles denote the single-step transition for $\tau = 0.5$. In panel (b), red filled circles denote the multi-step transition for $\tau = 0.5$, while blue filled circles denote the single-step transition for $\tau = 0$.}
	\label{1p2}
\end{figure}
\section{Dynamics of mean-field model}\label{analytical}
In this section, we employ the collective coordinate ansatz to analyze the synchronization dynamics of intrapopulation clusters~\cite{fialkowski2023heterogeneous,berner2023adaptive,smith2020model,smith2019chaos}. Within this framework, the high-dimensional oscillator dynamics is reduced to a low-dimensional description in terms of collective variables that capture the essential features of cluster evolution.

According to the collective coordinate ansatz, the phase variables $\phi_i^\mu$ and the coupling weights $k_{ij}^{\mu\nu}$ are approximated as
\begin{subequations}
	\label{an1}
	\begin{eqnarray}
		\phi_i(t) &\approx& \tilde{\phi}i^\mu (t) = \Theta\mu(t) (\omega_i - \Omega_\mu) + f_\mu(t),
	\end{eqnarray}
	\begin{eqnarray}
		k_{ij}(t) &\approx& \tilde{k}{ij}^{\mu\nu} = k{\mu\nu}(t),
	\end{eqnarray}
\end{subequations}
where the variable $f_\mu(t)$ represents the collective (mean) phase of the oscillators belonging to the $\mu$-th frequency cluster. The term $\Theta_\mu(t)(\omega_i - \Omega_\mu)$ describes the deviation of the $i$-th oscillator from the mean frequency $\Omega_\mu$ of cluster $\mu$. The microscopic coupling weights $k_{ij}$ are coarse-grained into effective cluster–cluster couplings $k_{\mu\nu}$, where $\mu,\nu \in {1,2}$ denote the cluster indices.
Substituting the ansatz given by Eqs.\eqref{an1} into the original system Eqs.\eqref{eq1} results in a nonzero residual (error) vector $\mathbf{E}$ of dimension $N + N^2$, with components
$\mathbf{E} = \bigl( E_{\phi_1^1}, \dots, E_{\phi_N^M}, \, E_{\kappa_{1,1}^{1,1}}, \dots, E_{\kappa_{N,N}^{M,M}} \bigr).$

The error corresponding to the phase dynamics and the coupling dynamics is given by
\begin{eqnarray}
	\label{an2}
	E_{\mu,i}^{\phi}(t) &=& \dot\Theta_\mu(t)(\omega_i - \Omega_\mu) + \dot{f}_\mu(t) \\ \nonumber
	&-& \omega_i + \frac{\sigma}{N}\sum_{\nu}\sum_{j\in c_\nu} k_{\mu\nu}(t) \sin(\phi_i^\mu(t) - \phi_j^\nu(t - \tau)),
\end{eqnarray}
\begin{eqnarray}
	\label{an3}
	E_{\mu\nu, ij}^{k} = \dot k_{\mu\nu}(t) + \epsilon[k_{\mu\nu}(t) + \sin(\phi_i^\mu(t) - \phi_j^\nu(t - \tau) + \beta)].
\end{eqnarray}
To minimize the residual, we enforce orthogonality of the error vector $\mathbf{E}$ to the tangent space of the ansatz manifold defined by Eqs.~\eqref{an1}. Denoting the ansatz by $u = {\tilde{\phi}, \tilde{k}}$ and the collective coordinates by $\mathbf{c} = {\Theta_\mu, f_\mu, k_{\mu\nu}}$, we project the error onto the directions $\partial u / \partial \Theta_\mu$, $\partial u / \partial f_\mu$, and $\partial u / \partial k_{\mu\nu}$.

Projection onto $\partial u / \partial \Theta_\mu$ yields
\begin{multline}
	\label{an4}
	\dot\Theta_\mu \sum_{i \in c_\mu} (\omega_i - \Omega_\mu)^2 = \sum_{i \in c_\mu} (\omega_i - \Omega_\mu) \omega_i \\ -\frac{\sigma}{N}\sum_{i \in c_\mu} (\omega_i - \Omega_\mu)\sum_{\nu=1}^{M}\sum_{j \in c_\nu}k_{\mu\nu} \sin(\phi_i^\mu(t) - \phi_j^\nu(t - \tau)), 
\end{multline}
Similarly, projection onto $\partial u / \partial f_\mu$ and $\partial u / \partial k_{\mu\nu}$ leads to
\begin{align}
	\label{an5}
	\dot f_\mu &= \Omega_\mu - \frac{\sigma}{N_\mu N}
	\sum_{i \in c_\mu}\sum_{\nu}\sum_{j \in c_\nu}
	k_{\mu\nu}\sin\!\bigl(\phi_i^\mu(t) - \phi_j^\nu(t - \tau)\bigr), \\
	\dot{k}_{\mu\nu} &=
	-\epsilon\left[
	k_{\mu\nu}
	+ \frac{1}{N_\mu N_\nu}\sum_{i \in c_\mu}\sum_{j \in c_\nu}
	\sin\!\bigl(\phi_i^\mu(t) - \phi_j^\nu(t - \tau) + \beta\bigr)
	\right].
\end{align}
To further simplify these equations, we consider the continuum limit $N \rightarrow \infty$ where sums of the form $\frac{1}{N} \sum_i g(\omega_i) = \int g(\omega) \rho(\omega) d\omega$. The order parameter for a single cluster is satisfies $r_\mu = \frac{1}{N_\mu} \vert \sum_{j\in\mu} e^{i\phi_i^\mu}\vert$. 

Substituting the reduced phase $\phi^\mu = \Theta_\mu (\omega_i - \Omega_\mu) + f_\mu$, we obtain
\begin{eqnarray}
	\label{order1}
	r_\mu = \frac{1}{N_\mu} \left\vert \sum_{j\in \nu}e^{i\Theta_\mu (\omega_i - \Omega_\mu) + f_\mu} \right\vert.
\end{eqnarray}
\begin{figure}[!ht]
	\centering
	\includegraphics[width=0.5\textwidth]{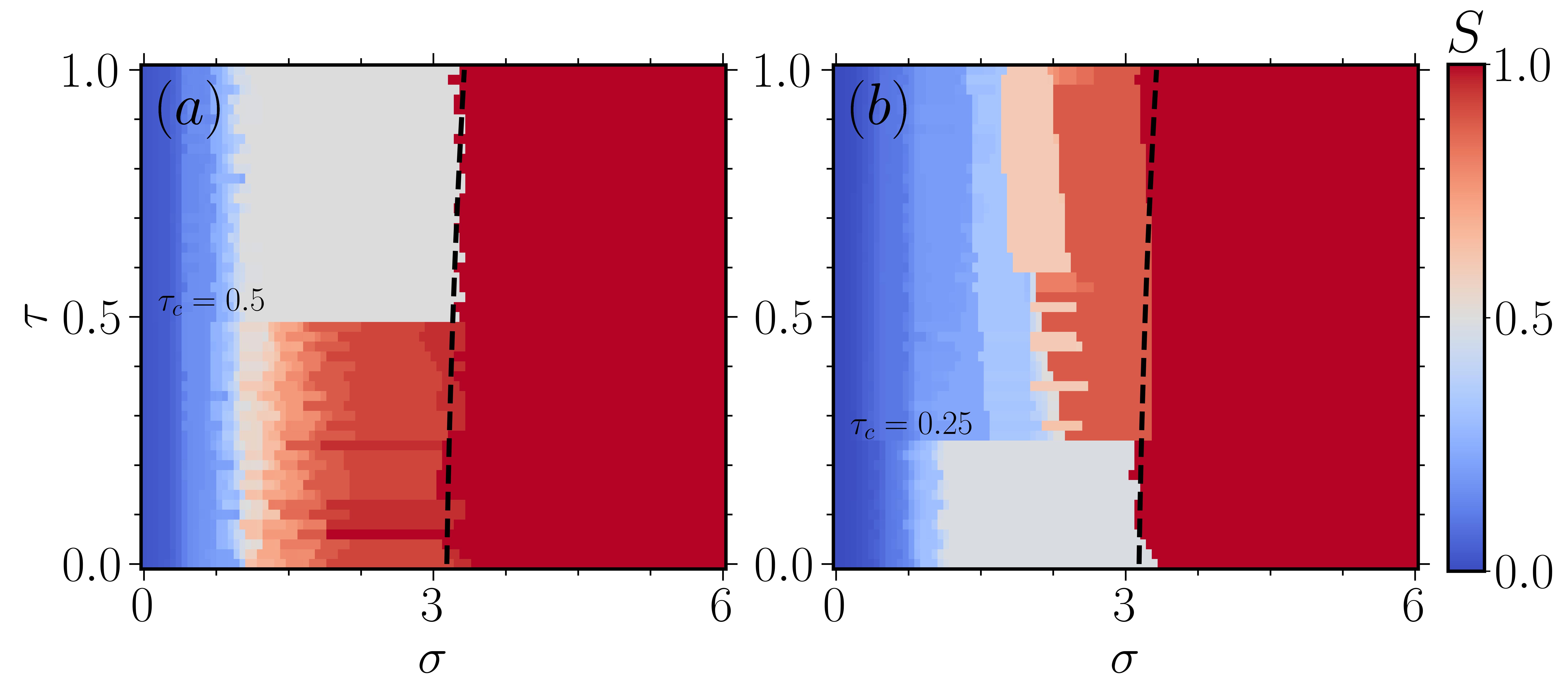}
	\caption{Phase diagrams of the synchronization index $S$ in the $(\sigma,\tau)$ parameter space, obtained from 10 different initial phase realizations. For each realization, $S$ is computed separately and plotted individually. The black dashed curve represents the analytical upper bound for the coupling strength. All other parameters are identical to those used in Fig.~\ref{1p_num}.}
	\label{2p}
\end{figure}
Since the order parameter measures phase coherence rather than absolute phase, the global phase factor $e^{if_\mu}$ does not affect its magnitude. Thus, we obtain
\begin{eqnarray}
	\label{order2}
	r_\mu = \frac{1}{N_\mu} \left\vert \sum_{j\in \nu}e^{i\Theta_\mu (\omega_i - \Omega_\mu)} \right\vert.
\end{eqnarray}
In the continuum limit, this expression becomes
\begin{eqnarray}
	r_\mu \approx \left| \int \rho_\mu(\omega) e^{i\Theta_\mu (\omega - \Omega_\mu)} d\omega \right|.
\end{eqnarray}

Assuming a uniform distribution within the cluster interval, $\rho_\mu(\omega) = 2/n_\mu$ for $\omega \in [\Omega_\mu - n_\mu/4, \Omega_\mu + n_\mu/4]$, we obtain
\begin{eqnarray}
	r_\mu = \frac{4}{n_\mu \Theta_\mu} \sin\left(\frac{n_\mu \Theta_\mu}{4}\right).
	\label{order3}
\end{eqnarray}
Substituting these expressions into the evolution equations yields the reduced system governing the collective coordinates:
\begin{eqnarray}
	\label{an6}
	\dot\Theta_\mu &=& 1 + \frac{\sigma}{v_\mu \Theta_\mu}\left[\cos\left(\frac{\Theta_\mu n_\mu}{4}\right) - r_\mu\right] \\
	&\times& \left[\sum_{\nu}n_\nu r_\nu k_{\mu\nu} \cos(f_\mu(t) - f_\nu(t-\tau))\right], \nonumber \\ \nonumber
	\dot f_\mu &=& 	\Omega_{\mu\nu} - r_\mu \sigma \sum_{\nu} n_\nu r_\nu k_{\mu\nu}	\sin(f_\mu(t) - f_\nu(t-\tau)), \\
	\dot k_{\mu\nu} &=& -\epsilon[k_{\mu\nu} + r_\mu r_\nu \sin(f_\mu(t) - f_\nu(t-\tau) + \beta)], \nonumber
\end{eqnarray}
Here, $\Omega_{\mu\nu} = \Omega_\mu - \Omega_\nu$, and $v_\mu$ denotes the variance of the natural frequencies within cluster $\mu$, defined as	$v_\mu = \frac{1}{N_\mu} \sum_{i \in \mathcal{C}\mu} (\omega_i - \Omega\mu)^2$.

In the continuum limit, the mean frequency and variance become \[
\Omega_\mu = \int \rho_\mu(\omega)\,\omega\,d\omega, \quad
v_\mu = \int \rho_\mu(\omega)(\omega - \Omega_\mu)^2 d\omega,
\]
which yield $\Omega_\mu = (n_\mu - 1)/4$ and $v_\mu = n_\mu^2/48$~\cite{smith2019chaos,fialkowski2023heterogeneous}.

The synchronization index  \eqref{S} becomes
\begin{eqnarray}
	\label{an7}
	S = \sum_{\mu} \sum_{\nu} n_\mu n_\nu s_{\mu\nu},
\end{eqnarray}
where $s_{\mu\nu} = 1$ if $\langle \dot{f}_\mu \rangle = \langle \dot{f}_\nu \rangle$, and $s_{\mu\nu} = 0$ otherwise.

The phase transitions $S$ described by \eqref{an7}, obtained from the evolution equations of the collective coordinates \eqref{an6}, are shown in Fig.~\ref{1p2} for both the absence of delay ($\tau = 0$) and the presence of delay ($\tau = 0.5$), corresponding to class-I and class-II frequency distributions, respectively. A prescribed six-cluster configuration is adopted, since a small number of clusters is sufficient to capture a single-step transition, whereas a larger cluster structure is needed to represent the intermediate multicluster states associated with the multi-step transition. For class-I distributions, the cluster mean frequencies are chosen as $\Omega_\mu = (-0.225,,-0.135,,-0.045,,0.045,,0.135,,0.225)$, while for class-II distributions they are taken as $\Omega_\mu = (-0.225,,-0.185,,-0.145,,0.145,,0.185,,0.225)$. These values are selected to emulate the effective class-I and class-II frequency organizations observed in the simulations.

For class-I frequency distributions [Fig.~\ref{1p2}(a)], the system exhibits multi-step synchronization transitions in the absence of delay ($\tau = 0$), whereas the introduction of delay ($\tau = 0.5$) leads to a transition to a single-step synchronization behavior. In contrast, for class-II frequency distributions [Fig.~\ref{1p2}(b)], the system displays single-step transitions when no delay is present, while multi-step transitions emerge in the presence of delay. In the figure, presence of delay is indicated by red filled circles, whereas absence of delay is denoted by blue filled circles. These results were corroborated with Fig.\ref{1p_num}.

We now proceed to derive explicitly the upper bound condition for the existence of two-cluster states from the reduced collective coordinate equations~\eqref{an6}. To analytically capture these effects, we employ a perturbative expansion with respect to the small parameter $\epsilon$~\cite{fialkowski2023heterogeneous}. Furthermore, we assume that the phase difference between the two clusters, defined as $f = f_\mu - f_\nu$, evolves linearly in time, corresponding to a relative rotational motion with a constant phase velocity $\Omega'$. This assumption motivates the following expansion of the collective coordinates in powers of $\epsilon$:

\begin{subequations}
	\label{an8}
	\begin{eqnarray}
		\Theta_\mu(t) &=& \Theta_\mu^{(0)} + \epsilon \Theta_\mu^{(1)}(t) + \mathcal{O}(\epsilon^2), \\
		k_{\mu\nu}(t) &=& k_{\mu\nu}^{(0)} + \epsilon k_{\mu\nu}^{(1)}(t) + \mathcal{O}(\epsilon^2), \\
		f(t) &=& \Omega't + \epsilon f^{(1)}(t) + \mathcal{O}(\epsilon^2).
	\end{eqnarray}
\end{subequations}
Substituting into \eqref{an6} leads to a quadratic equation in $\Omega'$. The inter-cluster dynamics is then described by
\begin{multline}
	\label{an9}
	\epsilon\dot f^{(1)}(t)
	=(\Omega_\mu - \Omega_\nu)-\Omega'
	-\epsilon\sigma \frac{(r_\mu^{(0)}r_\nu^{(0)})^2}{2\Omega'} \\
	\Big[
	n_2\big(\sin(2(\Omega't+\Omega_\nu\tau)+\beta)-\sin\beta\big) \\ 
	-n_1\big(\sin(2(\Omega't-\Omega_\mu\tau)-\beta)+\sin\beta\big)
	\Big].
\end{multline}
It can be used to estimate the upper bound for the existence of the two-cluster solution. Solving the quadratic equation for real valued $\Omega'$ implies that
\begin{eqnarray}
	\label{an10}
	(\Omega_\mu - \Omega_\nu)^2 \ge  -2 \epsilon \sigma (r_\mu^{(0)} r_\nu^{(0)})^2 \sin(\beta + \Omega_{\mu\nu} \tau).
\end{eqnarray}

Now, the upper bound of the coupling strength as a function of delay for the existence of the two-cluster state can be obtained as
\begin{eqnarray}
	\label{an11}
	\sigma_S(\tau) = \frac{(\Omega_\mu - \Omega_\nu)^2}{-2\epsilon(r_\mu^{(0)} r_\nu^{(0)})^2 \sin(\beta + \Omega_{\mu\nu} \tau)}.
\end{eqnarray}
Here, we take $r_\mu^{(0)} \approx r_\nu^{(0)} \approx 1$ and $\Omega_{\mu\nu} = \Omega_\mu - \Omega_\nu = -0.25$, while all other parameters are the same as in Fig.~\ref{1p_num}. The analytical condition shows strong agreement with the numerical results, as illustrated by the black dashed curve in the two-parameter phase diagram shown in Fig.~\ref{2p}.

Figure~\ref{2p}(a) presents a heat map of the synchronization index $S$ in the $(\sigma, \tau)$ parameter space for class-I frequency distributions. Similarly, Fig.~\ref{2p}(b) displays the corresponding two-parameter phase diagram for class-II frequency distributions. From both panels, it is evident that the time delay $\tau$ plays a crucial role in controlling the transition between different synchronization regimes. In particular, beyond a critical delay value $\tau_c$, the system undergoes a transition from one type of synchronization behavior to another. Notably, the value of $\tau_c$ depends on the nature of the underlying frequency distribution. We therefore proceed to analyze the scaling behavior of $\tau_c$ for both class-I and class-II distributions.

The critical delay $\tau_c$ at which cluster reorganization occurs can be estimated by the condition that the delay-induced phase shift becomes comparable to (i.e., of order unity with respect to) the intrinsic phase differences arising from frequency mismatches between clusters. This argument leads to the scaling relation~\cite{yeung1999time,montbrio2006time}
\begin{eqnarray}
	\label{tau_c}
	\tau_c \propto \frac{1}{|\Delta \Omega|}.
\end{eqnarray}
where $\Delta \Omega$ denotes the characteristic frequency difference between the emerging synchronized clusters. 

\begin{figure}[!ht]
	\centering
	\includegraphics[width=0.5\textwidth]{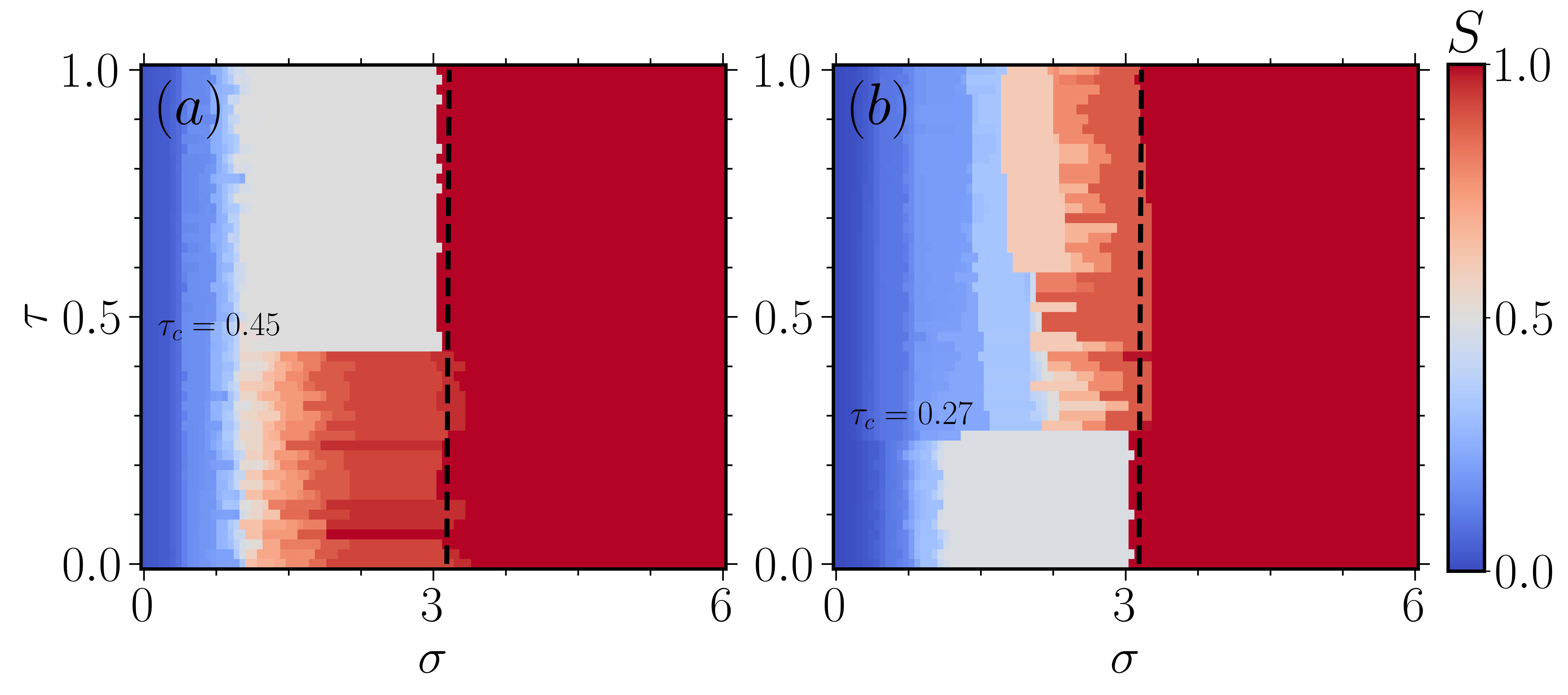}
	\caption{Phase diagrams of the synchronization index $S$ for the distributed-delay system \eqref{eq1_dd} in the $(\sigma,\tau)$ parameter space, obtained from 10 different initial phase realizations. For each realization, $S$ is computed separately and plotted individually. The black dashed curve represents the analytical upper bound for the coupling strength. All other parameters are identical to those used in Fig.~\ref{1p_num}.}
	\label{2p2}
\end{figure}
The critical delay $\tau_c$ is determined by the characteristic frequency separation between clusters. For class-I distributions, which are symmetric around zero, the mean frequency vanishes and therefore cannot be used to define a timescale. Instead, the relevant frequency scale is set by the typical deviation from the mean. For a uniform distribution of width $\omega$, this is given by the median distance from the mean, which scales as $\omega/4$. In contrast, class-II distributions consist of two symmetric clusters centered about zero, and the appropriate frequency scale is given by the separation between these clusters, which is of order $\omega/2$. Consequently, the effective frequency mismatch is larger for class-II distributions, leading to a smaller critical delay $\tau_c \sim 1/|\Delta \Omega|$ compared to the class-I case. 

As a consequence, the critical delay satisfies $\tau_c^{\text{(class-I)}} > \tau_c^{\text{(class-II)}}$. This follows from the fact that a larger characteristic frequency mismatch ($\Delta \Omega$) leads to a smaller critical delay required for the delay-induced phase lag to become significant enough to break cluster symmetry and induce reorganization.

\section{Effect of distributed delay} 
\label{dis}
In this section, we investigate the phase transition behavior of adaptively coupled phase oscillators in the presence of heterogeneous (distance-dependent) delays. In contrast to the uniform delay $\tau$, we introduce pair-specific delays $\tau_{ij}$. A uniform transmission delay is an idealized assumption that may not hold in spatially distributed networks, where transmission times depend on the separation between oscillators. To examine whether the delay-controlled synchronization transitions reported above persist beyond this simplified setting, we extend our analysis to distance-dependent delays $\tau_{ij}$ for oscillators arranged on a ring. In this way, the distributed-delay formulation serves to test the robustness and generality of the proposed mechanism under a more realistic delay structure. Accordingly, the original system given in Eq.~\eqref{eq1} is modified as follows~\cite{timms2014synchronization,zanette2000propagating}:
\begin{subequations}
\label{eq1_dd}
\begin{eqnarray}
\label{eq1a_dd}
\dot\phi_i(t) &=& \omega_i - \frac{\sigma}{N} \sum_{j=1}^{N} k_{ij} \sin\!\bigl(\phi_i(t) - \phi_j(t - \tau_{ij})\bigr), \\
\label{eq1b_dd}
\dot k_{ij}(t) &=& -\epsilon \left[k_{ij} + \sin\!\bigl(\phi_i(t) - \phi_j(t - \tau_{ij}) + \beta\bigr)\right].
\end{eqnarray}
\end{subequations}
Here, the delay $\tau_{ij} = (\tau/N)\, d_{ij}$ between oscillators $i$ and $j$ increases linearly with their distance $d_{ij}$ under periodic boundary conditions on a ring. The distance-dependent delay is defined as
\begin{eqnarray}
	\label{dd}
	\tau_{ij} = \frac{\tau}{N} \min(|i-j|, N - |i-j|).
\end{eqnarray}
All other parameters and variables remain identical to those defined in Eq.~\eqref{eq1}.

We numerically integrate the modified system~\eqref{eq1_dd} using the same integration scheme described earlier. The simulations reveal the occurrence of both single-step and multi-step synchronization transitions for class-I as well as class-II frequency distributions, in both the absence of delay ($\tau = 0$) and the presence of delay ($\tau = 0.5$).

With the introduction of distributed delays $\tau_{ij}$, the reduced collective coordinate equations~\eqref{an6} must be reformulated as
\begin{eqnarray}
	\label{an12}
	\dot\Theta_\mu &=& 1 + \frac{\sigma}{v_\mu \Theta_\mu}\left[\cos\left(\frac{\Theta_\mu n_\mu}{4}\right) - r_\mu\right] \\
	&\times& \sigma \left[\sum_{\nu}n_\nu r_\nu k_{\mu\nu} \cos(f_\mu(t) - f_\nu(t-\tau_{\mu\nu}))\right], \nonumber \\ \nonumber
	\dot f_\mu &=& 	\Omega_{\mu\nu} - r_\mu \sigma \sum_{\nu} n_\nu r_\nu k_{\mu\nu}	\sin(f_\mu(t) - f_\nu(t-\tau_{\mu\nu})), \\
	\dot k_{\mu\nu} &=& -\epsilon[k_{\mu\nu} + r_\mu r_\nu \sin(f_\mu(t) - f_\nu(t-\tau_{\mu\nu}) + \beta)], \nonumber
\end{eqnarray}
In this reduced description, $\tau_{\mu\nu}$ represents the effective (average) delay between clusters $\mu$ and $\nu$, defined as the ensemble average over all oscillator pairs $(i,j)$ with $i \in \mu$ and $j \in \nu$, i.e., $\tau_{\mu\nu} = \langle \tau_{ij} \rangle_{i \in \mu,\, j \in \nu}$. 

Assuming a distance-dependent delay proportional to the inter-oscillator distance, $\tau_{ij} \propto d_{ij}$, where $d_{ij}$ is defined on a ring of size $N$, the possible distances are $d = 0, 1, \dots, \lfloor N/2 \rfloor$. For $d = 1, \dots, \lfloor N/2 \rfloor - 1$, there are $2N$ ordered pairs $(i,j)$ corresponding to each distance (accounting for both directions along the ring), while for the maximum distance $d = N/2$ (when $N$ is even), there are $N$ such pairs. 

Averaging over all ordered pairs $(i,j)$ with $i \neq j$, the mean distance is $\langle d_{ij} \rangle = N/4$. Consequently, the effective inter-cluster delay in the reduced model becomes
\[
\tau_{\mu\nu} = \frac{\tau}{4}.
\]

Using this result, the upper bound on the coupling strength for the existence of two-cluster states is given by
\begin{eqnarray}
	\label{an11}
	\sigma_S(\tau) = \frac{(\Omega_\mu - \Omega_\nu)^2}{-2\epsilon(r_\mu^{(0)} r_\nu^{(0)})^2 \sin\!\bigl(\beta + \Omega_{\mu\nu} \tau/4\bigr)}.
\end{eqnarray}

Here, $r_\mu^{(0)} \approx r_\nu^{(0)} \approx 1$ and $\Omega_{\mu\nu} = \Omega_\mu - \Omega_\nu = -0.25$, while all other parameters are the same as in Fig.~\ref{1p_num}. This analytical prediction is in strong agreement with numerical results, as illustrated by the black dashed curve in the two-parameter phase diagram shown in Fig.~\ref{2p2}.

Figure~\ref{2p2}(a) presents a heat map of the synchronization index $S$ in the $(\sigma, \tau)$ parameter space for the system~\eqref{eq1_dd} with class-I frequency distributions. Similarly, Fig.~\ref{2p2}(b) shows the corresponding phase diagram for class-II frequency distributions. From both panels, it is evident that the distributed delay plays a crucial role in controlling transitions between different synchronization regimes. In particular, beyond a critical delay value $\tau_c$, the system undergoes a transition from one synchronization state to another.

\section{Conclusion} \label{conclusion}
We have uncovered a robust and previously uncharacterized mechanism governing heterogeneous nucleation transitions in adaptively coupled Kuramoto networks, driven by the interplay of connection delays and intrinsic frequency heterogeneity. Our results establish the existence of two fundamentally distinct routes to global synchronization, abrupt single-step and hierarchical multi-step transitions, emerging from the coupled dynamics of adaptive interactions and finite-size frequency structure~\cite{fialkowski2023heterogeneous}.

A central outcome of this work is the identification of connection delay as a decisive control parameter that selects the synchronization pathway. Remarkably, this control is not universal but reverses with the underlying frequency distribution. For unimodal-like (class-I) distributions, delay suppresses hierarchical buildup and induces a direct, first-order transition to synchrony. In contrast, for bimodal-like (class-II) distributions, delay destabilizes abrupt synchronization and promotes the emergence of intermediate cluster states, leading to multi-step transitions. This reversal highlights a nontrivial and highly sensitive coupling between temporal delays and spectral properties of the oscillator population.

We further demonstrate that dynamically formed clusters acquire distinct collective frequencies, generating large phase velocity mismatches that far exceed those of individual oscillators. This emergent frequency separation acts as a dynamical barrier that inhibits inter-cluster synchronization over an extended coupling regime, thereby precipitating explosive, first-order transitions at critical coupling strengths~\cite{skardal2014complex,zhang2015explosive}. These findings provide a clear mechanistic explanation for abrupt synchronization in adaptive networks.
\begin{figure}[!ht]
	\centering
	\includegraphics[width=0.5\textwidth]{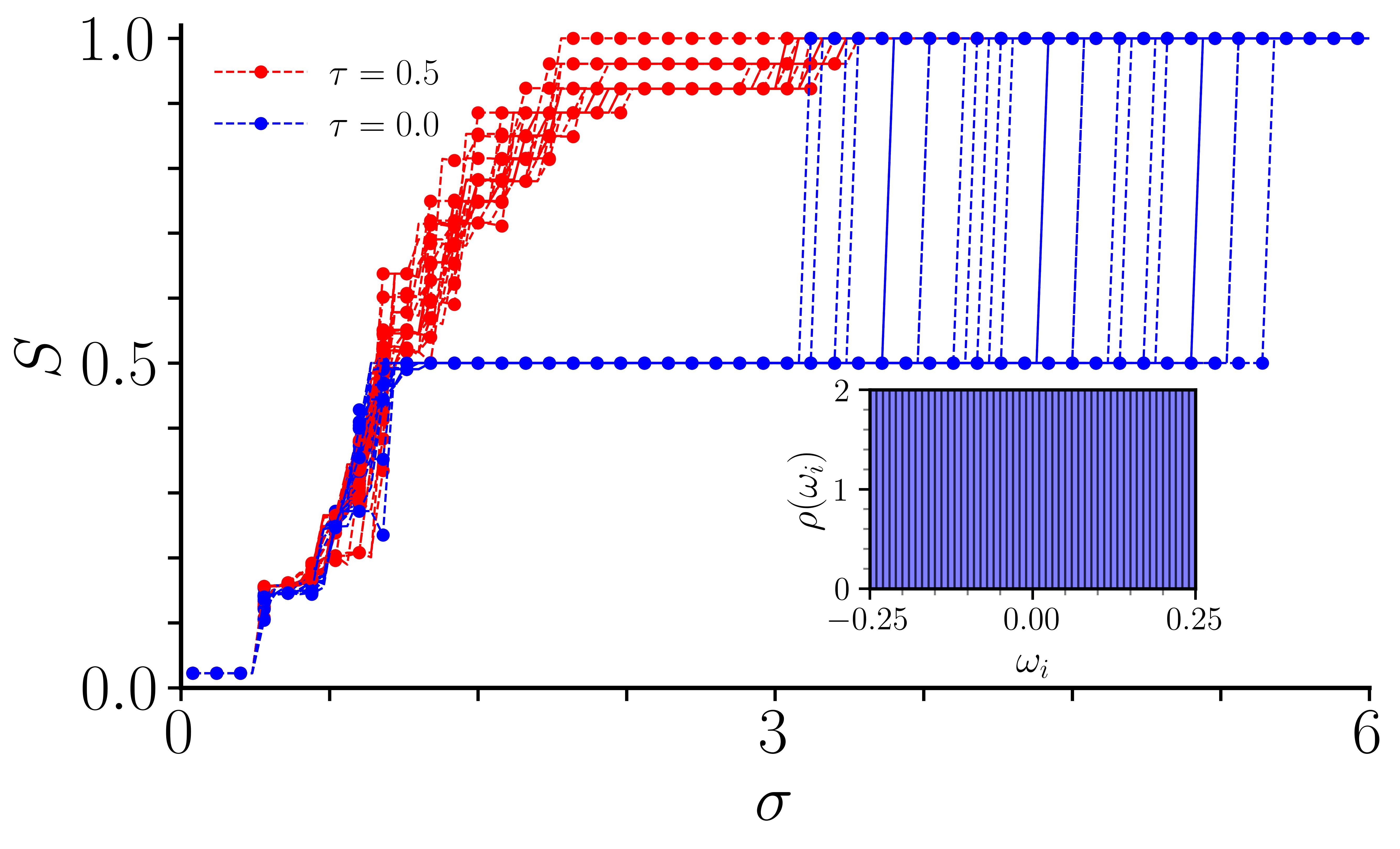}
	\caption{Synchronization index $S$ as a function of the coupling strength $\sigma$. Results are obtained from 100 independent simulations with equally spaced natural frequencies uniformly distributed in the interval $\omega_i \in [-0.25,\, 0.25]$. Blue filled circles denote the single-step transition observed for $\tau = 0$, while red filled circles indicate the multi-step transition for $\tau = 0.5$. The inset shows the probability density function of the natural frequency distribution.}
	\label{1p_uni}
\end{figure}

On the theoretical side, we developed a reduced macroscopic description using the collective coordinate framework~\cite{smith2020model,smith2019chaos}, incorporating both instantaneous and delayed interactions. The reduced equations quantitatively reproduce the full system dynamics and yield an explicit analytical upper bound for the coupling strength as a function of delay. This agreement establishes the collective coordinate approach as a powerful predictive tool for adaptive oscillator networks. At the same time, our results expose the current limitations of mean-field reductions in capturing the full complexity of multi-step nucleation, pointing to an important open direction for future theoretical development.

Finally, we generalized the framework to include distributed, distance-dependent delays~\cite{timms2014synchronization,zanette2000propagating}, demonstrating that the core phenomenology persists under more realistic interaction structures. These results open several promising avenues, including the role of higher-order interactions, directed or asymmetric coupling, and applications to biological and neural systems where delays and adaptation are intrinsic. 

Taken together, this work establishes delay–frequency interplay as a fundamental organizing principle of synchronization transitions in adaptive networks, and provides a unifying framework for understanding and controlling complex collective dynamics in systems with evolving interactions.

\appendix
\section{Uniform natural frequency with delay}
In this Appendix, we investigate the emergence of single-step and multi-step nucleation transitions arising from finite-size frequency inhomogeneity and connection delay. To this end, we numerically simulate the adaptive Kuramoto model described by Eqs.~\eqref{eq1}, employing an equally spaced uniform distribution of natural frequencies in the interval $\omega_i \in [-0.25,\, 0.25]$. Specifically, the frequencies are assigned as
\[
\omega_i = \frac{0.5(i-1)}{N-1} - 0.25,
\]
for a system of $N = 50$ oscillators.

Figure~\ref{1p_uni} summarizes the results from 100 independent realizations, each initialized with different random phase configurations. The simulations reveal a pronounced degree of multistability: as the coupling strength $\sigma$ is gradually increased, the system converges to distinct cluster configurations depending sensitively on the initial conditions.

In the absence of connection delay ($\tau = 0$), the system consistently exhibits a sharp, single-step transition to global synchronization. In contrast, when a finite delay is introduced ($\tau = 0.5$), the synchronization pathway becomes significantly more intricate, displaying clear multi-step nucleation characterized by intermediate partially synchronized cluster states.

Moreover, both with and without delay, small variations in the realization of the natural frequencies, together with differences in initial conditions, lead to qualitatively different synchronization routes. This highlights the combined role of finite-size effects and initial condition sensitivity in shaping the observed transition dynamics.

\section*{Acknowledgement}
R. A. acknowledges SASTRA for providing Teaching Assistantship. The research contributions of R. S. is part of a project funded by the SERB–CRG(Grant No. CRG/2022/004784). V. K. C. is supported by the ANRF Project under Grant No. ANRF/ARG/2025/004108/PS. The authors gratefully acknowledge the Department of Science and Technology
(DST), New Delhi, for providing computational facilities through the DST–FIST program under project number
SR/FST/PS-1/2020/135, awarded to the Department of Physics.  The work of J.F. was funded in part by the Austrian Science Fund (FWF) under grants 10.55776/I5985; and 10.55776/P34994.

\section*{DATA AVAILABILITY}
The data that support the findings of this study are available from the corresponding author upon reasonable request.

\bibliography{ref}
\end{document}